\documentclass[pra,notitlepage,nofootinbib,12pt]{revtex4-1}

\usepackage{bm}

\begin{document}

\title{Time-Symmetric Boundary Conditions and Quantum Foundations}

\author{Ken Wharton}

\affiliation{Department of Physics and Astronomy, San Jos\'{e} State University, San Jos\'{e}, CA 95192-0106}

\begin{abstract}

Despite the widely-held premise that initial boundary conditions (BCs)
corresponding to measurements/interactions can fully specify a physical subsystem, a literal reading
of Hamilton's principle would imply that both initial and final BCs are required (or more generally, a BC on a closed hypersurface in spacetime).
Such a time-symmetric perspective of BCs, as applied to classical fields, leads to interesting parallels with quantum theory.
This paper will map out some of the consequences of this counter-intuitive premise, as applied to covariant classical fields.  The most notable result is the contextuality of fields constrained in this manner, naturally bypassing the usual arguments against so-called ``realistic'' interpretations of quantum phenomena.

\end{abstract}



\maketitle

\section{Introduction}

Without measurement theory, theoretical physics would be a purely mathematical endeavor; it is the link between mathematical expressions and physical events that provides a testable connection between equations and reality.  But despite the temporal symmetries evident in our most fundamental physics equations, such symmetries are rarely applied to measurement theory itself.  Given a physical subsystem described by some equation, it is usually assumed that to find the solution to the equation one must impose initial boundary conditions (IBCs), but not final boundary conditions (FBCs).  This intuitive assumption leads to an asymmetric map between physical measurements and mathematical boundary conditions (BCs), where the map from the initial measurement to the IBC is treated differently than the map from the equation solution to the final measurement.  This asymmetry is especially evident in the realm of quantum theory.

It is notable that outside of measurement theory there exists a well-known, time-symmetrical application of mathematical boundary conditions: the use of Hamilton's principle to construct equations of motion from a Lagrangian \cite{Lanczos}.  In this case, one mathematically fixes both IBCs and FBCs (in coordinate space), and then determines the equations of motion for which the action is extremized.  Despite this symmetry, when using the resulting Euler-Lagrange equations to describe physical systems, one typically reverts to the usual procedure of imposing only IBCs (determined via measurements/interactions), ignoring the FBCs that generated the equations in the first place.  Presumably, in the minds of many physicists, there is an important distinction between mathematically imposing an FBC for the purpose of Hamilton's principle and actually imposing an FBC via a physical measurement.  However, there seems to be little distinction between these two where IBCs are concerned.

The purpose of this paper is to motivate and explore the consequences of treating measurement-BCs in the same symmetrical manner as they are treated in Hamilton's principle.  External measurements (both before and after the subsystem in question) will be treated as physical constraints, imposed on the subsystem in exactly the same way that boundaries are imposed when using Hamilton's principle.  This interpretation of measurements as BCs on the prior system (as opposed to merely revealing the status of events determined by IBCs) is counter-intuitive, but there is further motivation for such a framework (as seen in the next section).  The subsequent analysis indicates that classical fields constrained by FBCs exhibit many of the characteristics of quantum systems.  These results point to a natural resolution of several foundational issues in quantum mechanics, presented in a classical context.

\section{Motivating FBCs}

Variational principles notwithstanding, most physicists have a strong intuition that any given subsystem in the universe should be determined by IBCs alone.  Such an intuition is bolstered by the time-asymmetric second law of thermodynamics, which of course can only be derived using time-asymmetric assumptions.  The most natural such assumption is that of correlated IBCs (and no corresponding FBCs), from which coarse-grained entropy can be statistically shown to increase \cite{Schulman}.  Still, far from ``proving'' that nature is fundamentally time-asymmetric, all this demonstrates is that our observable universe happens to have an IBC which is highly correlated on a coarse-grained level (the low-entropy Big Bang).

Furthermore, when one considers single- or few-particle phenomena,
it is not obvious that lessons from statistical regimes should apply
in this more fundamental context.  Indeed, if the quantum level of
description is considered to be fundamental (fine-grained, with no
underlying level on which statistics can be applied), then one would
not expect any entropy-like asymmetry to arise from the use of
time-symmetric equations.  As Eddington pointed out long ago, ``When
[Dirac's theory] is applied to four particles alone in the universe,
the analysis very properly brings out the fact that in such a system
there could be no steady one way direction of time, and vagaries
would occur which are guarded against in our actual universe . . .
of about $10^{79}$ particles.''\cite{Eddington}  Recovering a
distinction between IBCs and FBCs on this fundamental level
therefore requires some new, time-asymmetric addition to the
equations (as Eddington also advocated).  But
time-symmetry is an
important principle in physics, so violating it in any way should
not be done lightly -- particularly when the phenomena being
described have no apparent time-asymmetry to explain.

Although time-symmetry and Hamilton's principle are two reasons for considering FBCs on the same footing as IBCs, another compelling reason is the experimental fact of the Heisenberg uncertainty principle.  In the context that Heisenberg first envisioned \cite{UP}, this is not necessarily an indication of some difference between quantum and classical systems, but rather a strict limit on how accurately one can measure \textit{any} system.  Simply put, this principle makes IBCs based on actual initial measurements \textit{inherently insufficient} to describe the evolution of classical systems.  Without such a restriction, imposing FBCs would be trivial: together the IBCs and FBCs would overconstrain the system, and FBCs would not ``determine'' anything at all; they would merely reveal what was already deducible from the IBCs.  But there is, in fact, no experimental system with sufficient IBCs to make FBCs redundant.  Viewed in this light, the uncertainty principle seems to discourage total reliance on IBCs (although quantum theory has resisted this conclusion, as discussed in the next section).

The uncertainty principle does not merely imply that IBCs contain
insufficient data; it also generally indicates that IBCs will be
insufficient by a factor of two (in full concordance with both
time-symmetry and Hamilton's principle).  In the simplest possible
case, a single particle in a known potential requires an initial
position and an initial velocity; the uncertainty principle allows
exactly half of this information to be measured and imposed as an
IBC.  More generally, quantum measurement theory for a single
spinless particle permits measurements that constrain the initial
values of some complex scalar function $\psi(\bm{x},t_0)$ (to within
a global phase).  Such an IBC is sufficient to solve the
Schr\"odinger equation (SE), as the SE is a first-order differential
equation in time.  But the corresponding classical system, a complex
scalar field $\phi(\bm{x},t)$, is governed by the
second-order-in-time Klein-Gordon equation \cite{Goldstein}, and
requires double the initial data: $\phi(\bm{x},t_0)$ and
$\dot{\phi}(\bm{x},t_0)$.  Again, one is limited to half of the IBCs
needed for a solution to the corresponding classical problem
\cite{Wharton}.

Even more generally, the IBCs and FBCs can together be treated as a
single closed-hypersurface boundary condition in spacetime, as is
generally done in covariant extensions of Hamilton's principle.  In
this case, while one can no longer pick out a well-defined ``IBC"
(portions of the boundaries can be time-like, and therefore neither
IBCs nor FBCs), the same half-data restriction still seems to
generally hold locally on the boundary surface.  For example, the
time-like boundary of a perfect conductor in classical
electromagnetism can constrain 3 of the 6 components of the E and B
fields on the boundary, but leaves the other 3 components fully
unconstrained.  Further evidence that nature has provided us with
systems underconstrained by their IBCs comes from the
``$\psi$-epistemic'' view of quantum states espoused by Spekkens
\cite{Spekkens}.  In a simple toy theory, Spekkens demonstrates that
restricting ones initial knowledge of a classical system (by exactly
a factor of two!) leads to a vast array of consequences normally
associated with quantum phenomena, including interference, quantum
teleportation, and an apparent ``collapse'' upon measurement.  The
primary quantum phenomena \textit{not} recoverable from such a
picture are Bell-inequality violations.

But even this exception implies that one should seriously consider FBCs, something Spekkens's model does not do.  In his model, while the known IBCs are insufficient, the unconstrained parameters are assumed to be determined by some additional, \textit{random} IBC, yielding a well-defined probability distribution for future measurements of the corresponding unknown parameters.  This, of course, is the type of ``hidden variable" that was considered by Bell in the proof of his inequality \cite{Bell}, so it is no surprise that Spekkens's toy theory is also constrained in a similar manner.  Both Spekkens and Bell assumed that any future measurement on the system would merely \textit{reveal} what was already determined by (known or unknown) IBCs.

And yet experiments tell us that nature does not respect Bell's inequality, leading many physicists to consider dramatic changes to our best models of spacetime, introducing non-locality and/or preferred reference frames.  An alternate approach, much less popular but still scattered throughout the literature \cite{OCB,CWR,Cramer,Sutherland1,Price,Miller,Wharton1}, is to permit hidden variables that are somehow correlated with the future measurement settings, in which case violation of Bell-type inequalities becomes trivial.  The question of \textit{how} such retrocausal correlations might arise is not always specified, but the above analysis should point to an obvious option: perhaps the future measurement is not merely \textit{revealing} information, but actually a  \textit{constraint} on the prior development of the system.  Systems constrained by FBCs are naturally ``contextual'', in that their internal parameters \textit{depend} upon that future measurement/interaction.  One \textit{expects} Bell-inequality violations in such a system.  Perhaps the strongest reason to consider FBCs is that it might provide a way to explain such violations in a way compatible with the local nature of spacetime assumed by general relativity.

\section{The Case for Classical Fields}

The previous sections motivate an equivalence between external measurements on a spacetime subsystem and the mathematical boundary conditions on the underlying equations that describe that system.  Note that any connected region of spacetime only has one boundary on which external measurements can be imposed; if this region is effectively infinite in spatial extent, then the boundary is often considered to be formed by two spacelike surfaces.  So for any given region, three or more complete external measurements on the same subsystem are not logically possible.  (Measurement devices take up a spacetime volume themselves.)  This approach assumes that measurements made by physical devices are both constraints on spacetime-adjacent subsystems as well as being part of larger subsystems that are themselves constrained by other measurements/BCs.  In this view the ultimate boundary conditions are cosmological.

Given this framework where all measurements are BCs on some subsystem, there are three general categories of systems/equations that one might consider; classical particles, classical fields, and quantum fields.  Previous applications of FBCs to physical systems have primarily focused on classical particles \cite{Schulman} and quantum fields \cite{Schulman, Miller1, AV, Oeckl}.  (Sutherland recently pursued this idea in the context of Bohmian quantum mechanics, spanning both classical particles and quantum fields \cite{Sutherland}.)  Apart from my own research program (as well as recent work by Dolce \cite{Dolce}), there seems to be no one pursuing the application of FBCs onto classical fields.

Classical fields may seem like a curious middle ground between the other two options, more abstract than particles but not as foundational as quantum systems.  Still, there is a case to be made that classical fields constrained by closed hypersurface BCs might be an \textit{alternative} to quantum systems constrained by only IBCs.  First I will address the particle/field issue, and then tackle the classical/quantum question.

The strongest case for considering fields over particles is to look at our most successful fundamental physical theories, general relativity (GR) and quantum field theory (QFT).  GR is most naturally expressed in terms of fields (the Einstein equation is a field equation, derivable from a continuous Lagrangian density), and although it might be argued that this beautiful formalism is only an approximation of underlying discrete events, efforts to discretize spacetime itself have so far been unsuccessful.  QFT has even clearer field-based underpinnings, and analysis of inequivalent Fock-space representations in accelerating reference frames has demonstrated that ``particle number'' is not a generally covariant concept \cite{NoPs}.  This strongly implies that particles are not even ``real'', let alone fundamental.

Apart from these appeals to GR and QFT, it is notable that various efforts to describe quantum phenomena using a particle ontology (deBroglie-Bohm-style interpretations, stochastic electrodynamics, etc.) all are forced to introduce fields into the ontology as well as particles.  This is because it is extremely difficult to extract wave-like behavior from particles alone; conversely, it is trivial to describe a highly-localized field, at least at some particular instant.  For all of these reasons, it seems evident that fields should at least be given equal status to particles when asking foundational questions, if not strict priority.

When searching for answers to foundational questions, one may be tempted to not consider classical systems at all.  After all, it is widely known that classical physics fails in many instances, apparently forcing us into a quantum framework.  But there is a key flaw in this argument.  What is known to fail are classical systems \textit{constrained by IBCs alone}.  Such systems cannot possibly violate Bell-type inequalities, for example, so this implies is that some change must be made to the ``classical + IBC-only'' framework.  Modern physics has already made such a change, jumping to the ``quantum + IBC-only" framework that we have found so difficult to reconcile with GR.  While the above arguments may tempt one to go straight to a ``quantum + IBC + FBC'' framework (as is being pursured by Oeckl and others \cite{Oeckl}), such a path would bypass the mostly unexplored ``classical + IBC + FBC'' framework.  If this latter framework could address the experiments that motivate the classical $\to$ quantum transition, then \textit{there would be no need to ever leave the classical framework to begin with}.  This would provide a possibility to \textit{explain} the origin of some of the curious postulates of quantum theory, rather than just assuming them from the outset.

Indeed, the traditional classical $\to$ quantum transition has effectively bypassed some of the arguments for FBCs, making the addition of FBCs to the quantum framework an attempt to solve the same problem in two different ways.  The uncertainty principle, for example, is no longer seen as statement of unknown parameters that might be supplied by an FBC, but typically interpreted to mean that the system can still be perfectly well-defined using only half of the classical parameters.  Quantum theory accomplishes this halving of the parameter space by dropping from classical second-order (in time) partial differential equations to the first-order equations described by the SE and its natural relativistic extensions (e.g. the Wheeler-deWitt equation).  It is notable that this ``halving'' introduces severe problems in curved space-time, where there is no covariant way to separate out the underlying second-order equations into two first-order equations \cite{DeWitt}.  Viewed in this light, the entire structure of quantum theory is built around the premise that one should be able to do physics with IBCs only; adding FBCs to this framework is therefore quite unnatural.

This conclusion is evident in the results of the few researchers who have explored the ``quantum + IBC + FBC'' framework.  Schulman has found that imposing both IBCs and FBCs to the SE is a strong overconstraint, and has had to resort to finding approximate solutions that are necessarily inexact \cite{Schulman}.  Aharanov and Vaidman's two-state formalism \cite{AV} has overcome this problem by re-doubling the size of the quantum state to permit the addition of FBCs; a similar treatment can be found in other two-time boundary approaches \cite{Sutherland, Wharton1, Miller1}.  (Miller's approach \cite{Miller1} is arguably more similar to Schulman's, in that it resorts to non-unitary evolution to get from the IBC to the FBC.)  I have argued \cite{Wharton} that it is far more natural to start with a second-order partial differential equation (like the Klein-Gordon equation) and apply both IBCs and FBCs rather than to halve-and-then-double the parameter space to accomplish effectively the same thing -- especially given that this ``halving'' procedure fails in curved spacetime \cite{DeWitt}.

In this perspective, the ``classical field + IBC + FBC'' framework should be viewed as a competitor to the ``quantum + IBC'' framework; both are attempting to deal with limited IBCs in a completely different manner.  While it is certainly reasonable to have doubts that such a classical framework might explain the same experiments as quantum theory, it remains a fact that this framework has been largely unexplored.  The below analysis of the first-order consequences of the ``classical field + IBC + FBC'' framework does not encounter any major problems, and even supports its general plausibility.  Section 5 then discusses how many of the difficult interpretational questions of quantum theory would vanish under such a classical alternative.

\section{First-Order Consequences: Quantization and Contextuality}

For those unfamiliar with two-time boundary problems, the simplest and best analogy is that of two-spatial boundary problems.  Such systems are widely familiar in contexts ranging from high-Q laser cavities to a quantum particle in an infinite square well.  In both of these examples, the enclosed field is found to have a quantized wavelength (and therefore a quantized wavenumber) $k_n=n \pi/L$, where $L$ is the distance between the boundaries.  Notice that quantization is a natural consequence of two-boundary problems, even for classical systems (such as electromagnetic fields in laser cavities).

Applying this same two-boundary logic to a field subject to both an IBC (at time $t=t_0$) and an FBC (at time $t=t_f$), leads to a similar quantization of the ``temporal wavelength'', aka the period.  The corresponding ``wavenumber" in this case is just the angular frequency $\omega$, which should therefore be quantized in the same manner:
\begin{equation}
\label{eq:fquant} \omega_n=\frac{n\pi}{\Delta t}
\end{equation}
where $\Delta t = t_f - t_0$.  This is a novel ``frequency quantization'' which would seem to necessarily occur given both an IBC and FBC.

Sequential measurements of fields does not usually reveal the quantization from Eqn. (\ref{eq:fquant}), and this may seem to be reason to reject this approach.  But consider that for most systems it is impossible to externally constrain $\Delta t$ in the same way that one can externally constrain $L$.  In fact, the only experimental set-up where this is generally possible is that of a laser cavity, where the length is directly related to the time between interactions via $L=c\Delta t$.  In this special case, the above ``temporal quantization'' leads to $\omega_n=n\pi c/L$, which indeed is the quantization that is actually observed in this instance.  Of course, one traditionally comes to this conclusion by using the relation between the frequency and the quantized wavenumber $\omega_n=ck_n$, so this known result is \textit{consistent} with the use of FBCs, but not \textit{dependent} upon it.

For fields that do not propagate at some known speed, it is quite difficult to physically constrain $\Delta t$ to the accuracy where one might see the effects of (\ref{eq:fquant}).  For example, consider a classical scalar field $\phi$ governed by the generally covariant Lagrangian density
\begin{equation}
\label{eq:Lag} {\cal{L}}=\frac{1}{2} \left(
g^{\mu\nu}\nabla_\mu\nabla_\nu \phi - \frac{m^2c^2}{\hbar^2}\phi^2
\right)
\end{equation}
Here $\nabla$ is the covariant 4-derivative, summation is implied, and $g$ is the metric.  The parameters $m$ and $\hbar$ are chosen such that the corresponding Euler-Lagrange equation in flat Minkowski spacetime
\begin{equation}
\label{eq:ELE}
\left( \frac{\partial^2}{\partial t^2} -c^2\nabla^2 + \frac{m^2c^4}{\hbar^2}\right)\phi=0
\end{equation}
has the dispersion relation $\omega (\bm{k})=\sqrt{c^2 k^2+m^2c^4/\hbar^2}$.  This matches the energy-momentum relationship for a relativistic particle with mass $m$, subject to the deBroglie relations $E=\hbar\omega$, $\bm{p}=\hbar\bm{k}$.  Equation (\ref{eq:ELE}) is the Klein-Gordon equation (KGE), and of course is a second-order partial differential equation (in time).

Comparing the dispersion relation to (\ref{eq:fquant}), one finds
\begin{equation}
\label{eq:limit} \frac{n\pi}{\Delta t} = \omega_n \ge
\frac{mc^2}{\hbar}
\end{equation}
\begin{equation}
\label{eq:Compton} \Delta t \le n \pi \frac{\hbar}{mc^2}
\end{equation}
From this one can see see that this quantization requires the time between the IBC and FBC to be constrained to an accuracy less than the Compton period $h/(mc^2)$.  As this value is less than $10^{-20}$ seconds for electrons, and even smaller for more massive particles, it is unsurprising that the frequency quantization implied by (\ref{eq:fquant}) is not observed (outside of laser cavities).

However, our current inability to make measurements at well-defined sub-attosectond intervals does not mean that this framework has no consequences.  For example, consider two consecutive measurements made on this scalar field with the additional constraint of a 1D infinite-potential square well of length L.  Now there are two, independent, quantization conditions: the usual wavenumber quantization $k =n_x \pi/L$ and the new frequency quantization $\omega = n_t \pi/\Delta t$.  Inserting these expressions into the dispersion relation yields the constraint
\begin{equation}
\label{eq:ns} n_t^2 L^2 + n_x^2 c^2 (\Delta t)^2 =
\frac{m^2c^4}{\pi^2\hbar^2} L^2 (\Delta t)^2
\end{equation}

The fact that $n_x$ and $n_t$ are both integers is now a strong constraint on $\Delta t$ (given a well-defined value for L).  Most values of $\Delta t$ will have no solution at all, and for those values that do have solutions, the vast majority will now have only one integer pair ($n_x$,$n_t$) that satisfies (\ref{eq:ns}).  Therefore, given two consecutive measurements on a classical field in a 1D cavity, one expects to typically find well-defined single values of $k$ and $\omega$; \textit{not} the superposition of modes that one sees in a laser cavity where $n_x=n_t$.  In other words, particle-like measurement results have emerged from simply constraining classical fields with FBCs.  (Note that if no final measurement was made, there would have been no such constraint, and a generic superposition of solutions to the KGE would still be permitted.)  It seems plausible that this general result would continue hold for any potential well, not merely the infinite square well.

Another interesting result arises from the previous example: The time between measurements $\Delta t$ is no longer a completely arbitrary external parameter.  There are values of $\Delta t$ with no acceptable solution, and therefore measurements may not occur at certain times.  This is consistent with the probabilistic treatment of measurement time in \cite{Wharton}, but is certainly contrary to how most physicists think about this parameter.  The use of FBCs has now dictated which hypersurfaces may correspond to physical measurements, and which ones may not.  Taking this logic to its conclusion, one finds that such constraints on classical field measurements (as dictated by global solutions to an action-extremization problem) naturally leads to exactly the sort of quantization observed in experiments, including the quantization of angular momentum in units of $\hbar/2$ \cite{Wharton3}.

This relates to a technical issue concerning the boundaries in covariant formulations of Hamilton's principle.  Action-extremization only generally leads to the Euler-Lagrange equations if the boundaries constrain the value of the field in coordinate space \cite{Lanczos, Wharton3}.  If these boundaries are now to be interpreted as physical measurements, one encounters a problem: constraints on boundaries in coordinate space necessarily correspond to measurements of time-even quantities (parameters that are identical under time-reversal).  For example, the $j$-momentum density of a classical scalar field, as defined by the stress-energy tensor component $T^{0j}$, is proportional to $\dot{\phi}\,\partial\phi/\partial x^j$, a time-odd quantity.  Such a measurement cannot be imposed as a BC on $\phi$ in coordinate space, because $\dot{\phi}$ is independent from $\phi$ on any space-like hypersurface.  If one assumed complete experimental control over when and where measurements could be made, this would imply that Hamilton's principle would fail any time a time-odd quantity was measured.

Fortunately, the analysis in \cite{Wharton3} demonstrates that there is an interesting ``loophole'' that can save Hamilton's principle for time-odd measurements.  It turns out that boundary constraints on $\dot{\phi}$ can still permit an extremized action (leading to the same Euler-Lagrange equations), provided that the hypersurfaces corresponding to those boundaries have a particular geometrical structure.  (It is the constrained geometry of the hypersurfaces that leads to angular momentum quantization in \cite{Wharton3}.)  As above, this implies that our assumed freedom of when and where we can make measurements may be illusory, and indeed this lack of freedom seems to lead to an apparent measurement-induced quantization.

But the most important consequence of the above example is simply that the global solution to the equations in a space-time subsystem \textit{depend upon the next measurement} on that subsystem.  Even if we cannot control $\Delta t$ on a Compton period time-scale, we can certainly control it at much grosser levels, and in Eqn (\ref{eq:ns}) such external control will lead to different values of $n_t$ \textit{throughout} the spacetime volume bounded by the two sequential measurements.  In other words, if we are to treat $\phi$ as a ``real'' field, our future measurement decisions at $t\approx t_f$ will affect the reality of the field at times $t<<t_f$.  Such influence is usually termed ``retrocausality''.

Remarkably, the retrocausality that emerges from such a picture is of the least objectionable sort, as it only affects the so-called ``inaccessible past'' \cite{Price}, hidden parameters that cannot be determined until their ``future cause'' at $t_f$.  (If one was to make a measurement between $t_0$ and $t_f$, then that intermediate measurement would still be at $t_f$ by definition.)  Therefore no paradoxes can be constructed.  Furthermore, these are precisely the sort of counter-intutive hidden parameters needed to violate Bell's theorem in a natural way.  Such contextuality is not an added, \textit{ad hoc} feature of this framework, but an inevitable result of treating IBCs and FBCs on the same conceptual footing.

\section{Lessons for Quantum Foundations}

Quantum measurement theory typically comes with a built-in temporal asymmetry: complete initial measurements always correspond to a pure-state IBC, while complete final measurements are often made on mixed states and are not imposed as FBCs.  The (generalized) Born rule that gives the conditional probabilities of a final measurement outcome (conditioned upon the IBC and subsequent time evolution) is typically used in a time-asymmetric manner, in the sense that it is always conditioning on the past.  Although one might choose to condition on the future \textit{instead} of the past, and use the Born rule in reverse, one would find that the FBC becomes a pure state and the intermediate quantum state is no longer the same as it was in the past-conditioned case.  The Born rule therefore seems inconsistent with Hamilton's Principle, as the latter imposes \textit{both} IBCs and FBCs on the same system in a time-neutral manner.

In both the forward- and backward- application of Born's rule, despite the different intermediate mathematics, the joint probability between the past and future outcomes remains unchanged \cite{ABL}; it is this joint probability that therefore appears more fundamental than conditional probabilities when approaching such time-symmetric problems. For a complex classical field constrained by both an IBC and a FBC, I have shown that a invariant joint probability distribution can be constructed from the solution space of the intermediate field (given the boundary constraints) \cite{Wharton}.  From this one can construct the usual conditional probabilities in the appropriate limit, conditioning on any chosen portion of the boundary.  In this perspective, probability is not some mysterious attribute of quantum systems, but instead has the same source as it does in classical physics: ignorance.  Knowing the IBCs is not enough to define the fields, and therefore one retreats to a probabilistic framework.  Once the IBCs and FBCs are known, however, one can ``retrodict'' the field values in the intermediate region between the two boundaries if desired.

The so-called ``quantum collapse'' disappears in such a framework, in the same manner as envisioned by the ``$\psi$-epistemic'' proponents \cite{Spekkens}.  In this classical-field context, the solution to the SE ($\psi$) is merely our best guess of what the field looks like given only the IBC, but does not perfectly correspond to the actual field $\phi$ (which obeys the KGE, constrained by IBCs and FBCs).  Once the final measurement result (the FBC) becomes known, one updates ones knowledge of $\psi$ in a discontinuous manner.  Still, there is no discontinuity in reality, for the field $\phi$ is perfectly consistent with this final measurement result (as it is pre-constrained by the eventual FBC).

Another difference between $\psi$ and $\phi$ lies in the multi-particle sector.  To encode the probabilities of all possible measurements in a single function $\psi$, one is forced to expand the dimensionality of $\psi$ into multi-particle configuration space.  But in this framework the actual probabilities are not encoded by $\phi$ or $\psi$ themselves, but rather the solution space of $\phi$ given \textit{any particular future measurement set-up}.  There is no need for $\phi$ to encode the results of possible measurements which will not actually occur, so there is no need to expand $\phi$ into configuration space, even for multiple particles.  (Montina's recent work makes a related argument \cite{Montina}.)  The best example of this is the classical Maxwell field, which manages to reproduce infinite-photon-number quantum theory using fields that still exist in physical spacetime (as opposed  to infinite-dimensional configuration space).

The most important lesson from the above analysis is probably the following statement: Violations of Bell inequalities do not necessarily imply the failure of locality (in the sense used in general relativity).  Although locality is one assumption that underlies Bell's theorem, it is not the only one; the theorem also assumes that any hidden variables will not be dependent on the future measurement settings \cite{Bell}.  It is natural to violate such an assumption in this framework -- indeed, one \textit{expects} it to be violated, given a true FBC.  Experimentally-observed violations of Bell-type inequalities are therefore further motivation for considering FBCs, especially if one wants to retain basic notions of locality.

It is notable that this approach maps nicely to quantum field theory, which also uses an invariant expression for the joint probability of two consecutive measurements.  For a \textit{quantum} scalar field $\phi$ (promoted to an operator, with imposed commutation constraints), this joint probability is
\begin{equation}
\label{eq:qft} P[\phi(t_0),\phi(t_f)] = \left| \int {\cal{D}} \phi\,
e^{iS[\phi]/\hbar} \right| ^2
\end{equation}
Here the functional integral is over all field configurations consistent with the IBC $\phi(t_0)$ and the final measurement result $\phi(t_f)$; $S$ is the classical action.  It is notable that this mathematics naturally incorporates FBCs.  This expression can be made even more general by imposing the boundaries on a closed hypersurface rather than two parallel ``instants''.  Note that there is nothing, in principle, that would prevent the evaluation of (\ref{eq:qft}) for a purely classical field constrained by IBCs and FBCs.  In this perspective, the dominant contribution from quantum theory is on the boundary itself -- the limitation of the allowed measurement values on a given field.

But the classical framework presented here provides another way to constrain those measurements, without the use of quantum measurement theory (or operators of any sort).  If there is some pair of measurement results $\phi(t_0)$ and $\phi(t_f)$ with \textit{no} allowable solutions at all (such as certain values of $\Delta t$ and $L$ in (\ref{eq:ns})), then the integral in (\ref{eq:qft}) will simply be zero, and those pairs of measurements would therefore have zero probability of occurring together.  Furthermore, if there were pairs of possible measurements for which there was no way to extremize the classical action $S$ (such as angular momentum measurements that were not near a multiple of $\hbar/2$ \cite{Wharton3}),  then the value of the integral in (\ref{eq:qft}) would be quite small due to the lack of stationary phase terms.  The net effect would be an \textit{apparent} quantization of an underlying continuous system -- it would only be quantized when one ``looks", because it is the externally imposed observation/measurement that is the source of the quantization constraints.  Indeed, seeming paradoxes like ``delayed-choice'' experiments become trivial to interpret in this framework, as it is the future measurements themselves which are defining the evolution of the system.

The preliminary results presented here (and in related research \cite{Wharton,Wharton3}) are certainly nowhere close to being a replacement for all of quantum theory.  But there is every indication that this is a promising research avenue to pursue.  The only obvious conceptual disadvantage is the retrocausal aspect of FBCs, and even that might be considered an advantage if one is concerned about maintaining a temporal symmetry in theories that purport to describe time-symmetric events.  Indeed, taking Eqn. (\ref{eq:qft}) seriously can even be used as a strong case for retrocausation \cite{EPW,WMP}.  Furthermore, the framework advocated here is probably more compatible with general relativity than is quantum field theory, as classical fields in curved spacetime do not come with the conceptual problems of quantum fields.  It is even possible that, instead of quantizing spacetime, an approach to quantum gravity might be found in the continuous ontology of classical field theory.

For now, though, more modest efforts are called for: the construction of a time-neutral theory of field-field measurement, explicitly incorporating potentials/interactions, and searching for experimental tests of this overall framework.  Two-time boundaries imposed on classical fields is a subject that has been curiously neglected for the past hundred years.  Based on these preliminary results, it seems reasonable to venture further down this particular path.

\section{Acknowledgements}
The author is grateful to Rob Spekkens and the Perimeter Institute for the invitation to give the seminar on which this paper was based \cite{PItalk}.  Further thanks goes to Huw Price for conceptual inspiration and the Eddington quote.

\bibliographystyle{mdpi}
\makeatletter
\renewcommand\@biblabel[1]{#1. }
\makeatother

\end{document}